\documentclass[aps,aps,twocolumn]{revtex4}
\usepackage{graphicx}
\usepackage{dcolumn}
\usepackage{bm}
\usepackage{amsmath}
\usepackage{txfonts}
\usepackage[scaled]{helvet}
\usepackage{color}
\usepackage[breaklinks=true,colorlinks,citecolor=blue,linkcolor=blue,urlcolor=blue]{hyperref}
\usepackage{booktabs}
\usepackage{threeparttable}

\begin{document}

\title{Layered Opposite Rashba Spin-Orbit Coupling in Bilayer Graphene: Loss of Spin Chirality, Symmetry Breaking, and Topological Transition}
\author{Xuechao Zhai}
\affiliation{Department of Applied Physics and MIIT Key Laboratory of Semiconductor Microstructures and Quantum Sensing, Nanjing University of Science and
Technology, Nanjing 210094, China}

\date{\today}

\begin{abstract}
Inversion symmetry in bilayer graphene allows for layered opposite Rashba spin-orbit coupling (LO-RSOC) --- the situation when the RSOC has the same magnitude but
the opposite sign in two coupled spatially separated layers. We show that the LO-RSOC results in the loss of spin chirality in the momentum space, in contrast to the common uniform RSOC. This chirality loss makes it difficult to experimentally establish whether the LO-RSOC (on the scale of 10 meV) exists, because the band structure is insensitive to it. To solve this problem, we propose to identify the LO-RSOC either by gating to break the inversion symmetry or by magnetic field to break the time-reversal symmetry. Remarkably, we observe the transition between trivial and non-trivial band topology as the system deviates from the LO Rashba state. Ab inito calculations
suggest that bilayer graphene encapsulated by two monolayers of Au is a candidate to be a LO Rashba system.
\end{abstract}

\pacs{72.80.Vp, 73.43.Nq, 03.65.Vf, 75.70.Tj} \maketitle

\section{Introduction}

Monolayer graphene (MLG) has negligible spin-orbit couplings (SOCs), on the order of 10~$\mu{\rm eV}$ \cite{KonGmi,SiPra}, in the unperturbed state due to its high symmetries, typically inversion $(\cal I)$ symmetry, time-reversal $(\cal T)$ symmetry and $z\leftrightarrow-z$ (out of plane) mirror symmetry. Inducing SOC by symmetry breaking opens a distinctive route to explore the application of MLG in spintronics \cite{ManKoo,HanKaw,AvsOch}. As one of the most common SOCs, the Rashba SOC (RSOC) \cite{ManKoo} with the strength of more than 1~meV is relatively easily achieved in MLG by adatoms \cite{BalKok} or building hetero-interfaces, typically MLG$-$Au \cite{MarVar,FarTan}, due to $z\leftrightarrow-z$ mirror symmetry breaking. Interestingly, the RSOC in MLG induces an in-plane spin chirality manifest in spin-momentum locking and vortex-like spin polarization on the Fermi loop near each Dirac point \cite{Rashba}.

Bernal-stacked bilayer graphene (BLG) [Fig.~1(a)] consists of two MLG sheets, which are shifted by one bond length between each other and are weakly coupled by van der Waals (vdW) interaction \cite{McKosh}. Similarly to the MLG case, the SOCs in the natural-state BLG with $\cal I$ symmetry [see the black point for $\cal I$ center in Fig.~1(a)] and $\cal T$ symmetry are negligible \cite{McKosh,GelSmi}. Usually in theoretical model studies \cite{QiaoLi,MirSch,QiaoTse,GelSmi}, the RSOC in BLG is taken to be identical in sign or uniform for both layers, and in this situation it supports the in-plane spin chirality in the momentum space \cite{ZhaiJin}. In principle, this situation can be created, for example, by applying a vertical electric field, but this is highly inefficient because the estimated strength is only about $5~\mu{\rm eV}$ for a field of 1 V~nm$^{-1}$ \cite{HanKaw}. To date, the single-interface Rashba effect has been verified in BLG by putting it in proximity with transition-metal dichalcogenides \cite{WangChe,AlsAsm}, whereas the double-interface Rashba effect from both the top and bottom sides of BLG \cite{IsLew} is still not well understood and has been rarely explored.

Here we demonstrate that $\cal I$ symmetry in BLG allows for the situation when two layers have opposite RSOC---layered opposite (LO) RSOC. This results in the loss of spin chirality in the momentum space, in contrast to uniform RSOC. A crucial problem which follows is that, without symmetry breaking, it is hard to judge whether the LO-RSOC (on the order of 10~meV \cite{MarVar,BalKok}) exists or not since the band structure is insensitive to LO-RSOC. Here, we argue that the identification of LO-RSOC based on band structure becomes possible if there is at least one breaking for $\cal I$ symmetry and $\cal T$ symmetry. Remarkably, we demonstrate that there is a transition between trivial and non-trivial band topology, evidenced by the Berry's phase or the Chern number, when the system deviates from the LO-RSOC state. We further use {\it ab inito} calculations to show that BLG encapsulated by two monolayers of Au is a LO-RSOC system, for which the potential gradient along the $z$ axis to induce the LO-RSOC has opposite signs at the two opposite layers. The appearance of LO-RSOC here reveals a fundamental interaction phenomenon arising from symmetry. In contrast to the opening of the bandgap directly by LO Ising SOC \cite{IsLew} or by layered antiferromagnetism \cite{ZhaiXu} in doubly-proximtized BLG systems, the influence of LO-RSOC on electronic properties is highly hidden without symmetry breaking. Our results demonstrate the nontrivial effect of symmetry on spin properties and band topology.

Notably, the LO-RSOC discussed in BLG here provides a graphene-based version of Rashba bilayers, which have recently been widely explored in non-graphene systems, such as topological effects in quantum-tunneling-coupled Rashba bilayer heterostructures \cite{DasBal,RajBan,VolLoss}, hidden spin textures in Cu-based superconductors with two CuO layers \cite{Atkin,LuSen} or in covalently-coupled crystalline compounds \cite{LinWang,YuanLiu}, and chirality inversion on two opposite surfaces of 3D topological insulators \cite{XuXia,ChenKanou}. Compared with other Rashba bilayers, the vdW-coupled BLG combines many advantages of ultrathin materials, including simple structure, easy fabrication \cite{ZhouYu}, electrically-controllable high-mobility and band gap \cite{McKosh} and, most strikingly, being easy to assemble into a heterostructure \cite{IsLew,WangChe,AlsAsm}, making BLG especially attractive to experimentalists. Combined these factors with the possibility of miniaturization, BLG-based heterosystems are promising for exploration of spin-orbit physics and spintronics applications \cite{AvsOch,IsLew}. In contrast to the existing Rashba bilayers for which the band splitting is very sensitive to even weak perturbations induced by field-induced symmetry breaking  \cite{DasBal,RajBan,VolLoss,Atkin,LuSen,LinWang,YuanLiu,XuXia,ChenKanou}, the splitting for the Rashba BLG here is weakly sensitive to electrically or magnetically induced symmetry breaking that works within higher-order perturbations [Eq.~(\ref{dH})]. As BLG derives from the LO-RSOC state, the uniquely sharp topological transitions happen [see Eqs.~(\ref{BP}) and (\ref{Chern})]. These characteristics can be attributed to the specific vdW layered structure of BLG.

Our paper is organized as follows. In Sec.~II, we introduce the system Hamiltonian. In Sec.~III, we show the phenomenon and origin of chiral loss. In Sec.~IV, we demonstrate the field-induced symmetry breaking. In Sec.~V, we show the results of topological transition. Finally, we present the conclusions.

\section{System Hamiltonian}

\begin{figure}
\centerline{\includegraphics[width=8.5cm]{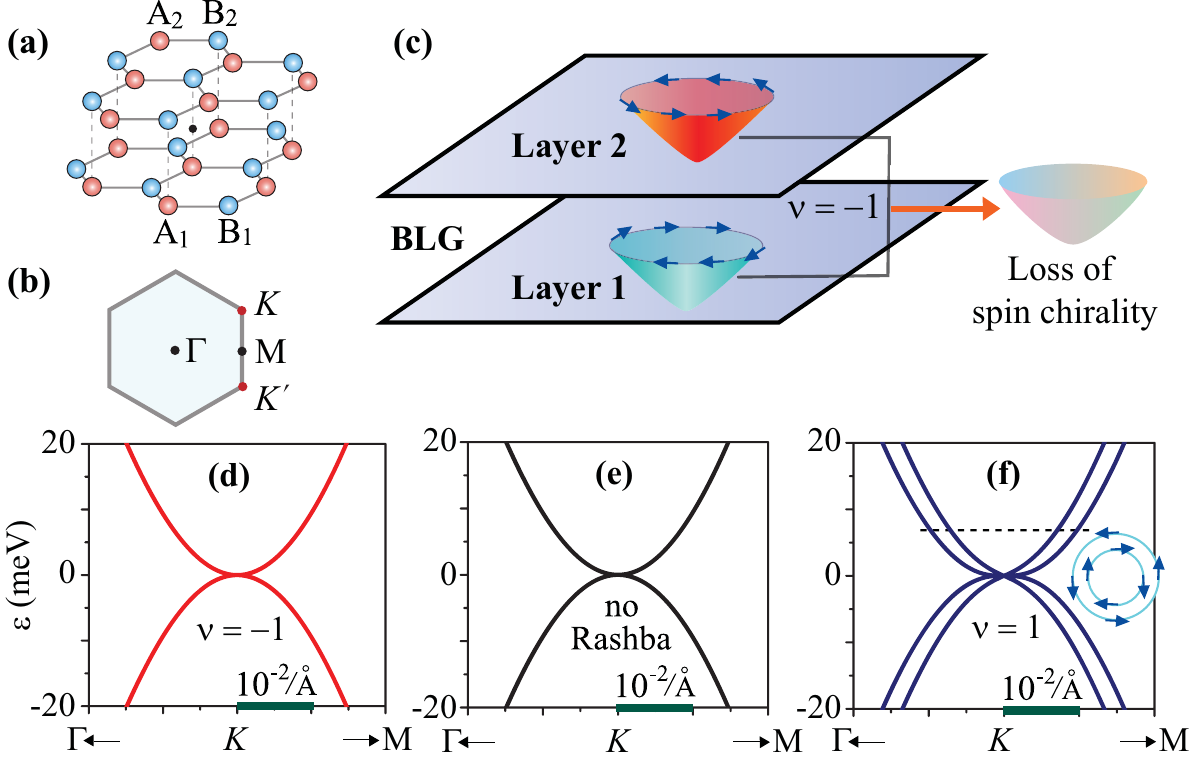}} \caption{(a) Bernal-stacked BLG. A$_{1(2)}$ and B$_{1(2)}$ denote two sublattices in carbon layer $1(2)$, and the black point marks an ${\cal I}$ center. (b) Brillouin zone. $\Gamma$, M, $K$ and $K'$ are four high-symmetry points. (c) Sketch of loss of spin chirality induced by LO-RSOC ($\nu=-1$). The Rashba-induced opposite spin chiralities in energy bands of two monolayers cancel out each other after vdW coupling. Blue arrows on each Fermi loop indicate the spin orientation. (d)-(f) Band structures for (d) LO-RSOC case ($\nu=-1$), (e) no Rashba case and (f) uniform-RSOC case ($\nu=1$). The inset in (f) plots the spin chirality on the Fermi loops at the dotted line.}
\end{figure}

According to the references \cite{McKosh,Neto,MarVar,BalKok,FarTan}, an empirical lattice Hamiltonian for Rashba BLG is constructed as follows
\begin{equation}\label{TBH}
\begin{split}
{\cal H}=&-t\sum_{{{\langle
i,j\rangle}_{\parallel}}\alpha}c_{i\alpha}^\dag
c_{j\alpha}-\gamma\sum_{{{\langle
i,j\rangle}_{\bot}}\alpha}c_{i\alpha}^\dag c_{j\alpha}\\
&+\frac{i\lambda}{3}\sum_{\langle i,j\rangle_{\parallel}\alpha\beta}
\chi_ic_{i\alpha}^\dag({\bm s}\times\hat{{\bm d}}_{ij})_{\alpha\beta}^zc_{j\beta}\\
&+U\sum_{i\alpha}\mu_ic_{i\alpha}^\dag c_{i\alpha}
+M\sum_{i\alpha}c_{i\alpha}^\dag s_zc_{i\alpha},
\end{split}
\end{equation}
where $c_{i\alpha}^\dag$ creates an electron with spin polarization $\alpha$ at site $i$, $\langle i,j\rangle$ runs over all the nearest-neighbor-hopping sites, and the subscript
${\|}~(\perp)$ means in-plane (out-of-plane), $\chi_i=1~(\nu)$ is valid when site $i$ is on the bottom (top) layer, $\mu_i=+1~(-1)$ holds if site $i$ locates on the bottom (top) layer, $\bm s$ is the spin Pauli operator, and $\hat{{\bm d}}_{ij}$ is the unit vector pointing from site $i$ to site $j$. There are five terms in total in Hamiltonian~(\ref{TBH}), and the parameters $t$, $\gamma$, $\lambda$, $U$, $M$ indicate the energy strength. The first and second terms represent the intralayer and interlayer nearest-neighboring hoppings, respectively. The third term denotes the Rashba SOC, which is not intrinsic in BLG but is inducible by interface engineering \cite{ManKoo} or adatoms \cite{BalKok} ($|\nu|\neq1$ essentially arises from $\cal I$ symmetry breaking in structure). The ratio of the Rashba coefficients of the top layer to the bottom layer is $\nu:1$, and hence $\nu$ can reflect the interlayer Rashba polarization, for which $\nu=-1~(+1)$ corresponds to the case of LO (uniform) RSOC. The fourth and fifth terms denote the other symmetry-breaking effects from gating ($2U$ is the vertical bias) that breaks ${\cal I}$ symmetry and magnetic field ($M$ is the Zeeman-splitting strength) that breaks ${\cal T}$ symmetry.

By performing the Fourier transformation \cite{McKosh,Neto}, a generalized eight-band Hamiltonian in the momentum space for Rashba BLG is derived as
\begin{equation}\label{H_EB}
\begin{split}
H(\bm p)=&\upsilon{\bm I_\tau}(\sigma_xp_x+\xi\sigma_yp_y){\bm I_s}+
\frac{\gamma}{2}(\tau_x\sigma_x-\tau_y\sigma_y){\bm I_s}\\
&+\frac{\lambda}{2}\tau_{z,\nu}(\sigma_xs_y-\xi\sigma_ys_x)
+U\tau_z{\bm I_\sigma}{\bm I_s}+M{\bm I_\tau}{\bm I_\sigma}s_z,
\end{split}
\end{equation}
which takes $\psi=\{\psi_{A_1\uparrow},\psi_{A_1\downarrow},\psi_{B_1\uparrow},
\psi_{B_1\downarrow},\psi_{A_2\uparrow},\psi_{A_2\downarrow},\psi_{B_2\uparrow},
\psi_{B_2\downarrow}\}$ as the atomic basis set. Here, ${\bm p}=(p_x, p_y)$ is used to denote the momentum by taking $K$ ($K'$) as coordinate origin. The Pauli matrices ${\bm s}=(s_x, s_y, s_z)$, ${\bm \sigma}=(\sigma_x, \sigma_y, \sigma_z)$, ${\bm \tau}=(\tau_x, \tau_y, \tau_z)$ are used to describe the spin, intralayer sublattice pseudospin and layer pseudospin degrees of freedom for electrons in BLG \cite{McKosh,GelSmi,QiaoTse,MirSch,QiaoLi,ZhaiJin}. The index $\xi=+1~(-1)$ marks valley $K$ ($K'$) in Fig.~1(b), and ${\bm I_s}$, ${\bm I_\sigma}$, ${\bm I_\tau}$ are used to label the identity matrix in the $\bm s,\bm\sigma$ and $\bm\tau$ spaces, respectively. Note that the five terms in Hamiltonian~(\ref{H_EB}) correspond to those in Hamiltonian~(\ref{TBH}) in order. Specifically, the first term $H_\upsilon$ in Hamiltonian~(\ref{H_EB}) indicates the massless Dirac term, where $\upsilon=\sqrt3at/2\hbar$ ($a=2.46~{\AA}$ is the lattice constant) is the Fermi velocity in MLG. In the third term, the $\nu$-dependent matrix in the $\bm \tau$ space reads
\begin{equation}
\tau_{z,\nu}\equiv
\left[\begin{array}{cc}
1&0\\
0&\nu\\
\end{array}\right],~~\nu\in[-1,1],
\end{equation}
which depicts the possible Rashba difference between two MLG sheets. Note that $|\nu|>1$ is not considered here because no more physics happens.

Below, the lattice Hamiltonian~(\ref{TBH}) is used for accurate band calculations. Unless otherwise noted ({\it e.g.} Fig.~2), the typical strength parameters $t=2.689$~eV, $\gamma=0.364$~eV and $\lambda=16.2$~meV (fit parameters extracted from Fig.~5) are used. Without doubt, the main Rashba physics we concern with does not change with the perturbation of parameters.

\section{Phenomenon and origin of chiral loss}

We consider the simplest case, $U=M=0$, in Hamiltonian~(\ref{H_EB}).
For MLG, the lowest-energy two subbands are derived as
$\varepsilon(\bm p)=\pm[(\lambda^2+4\epsilon^2)^{1/2}-\lambda]/2$ with  $\epsilon=\upsilon|\bm p|$. The average spin is derived as
$\langle{\bm s}\rangle=2{\rm sgn}(\lambda)\epsilon(\lambda^2+4\epsilon^2)^{-1/2}(\hat{{\bm e}}_{\bm p}\times\hat{\bm z})$ with $\hat{{\bm e}}_p=\bm p/|\bm p|$, and $\hat{\bm z}$ is the unit vector of the $z$ axis. Hence, the low-energy electrons possess spin chirality, as shown in Fig.~1(c), where opposite RSOC induces opposite spin chirality in opposite layers. As two monolayers gradually approach from an uncoupled state to a vdW-coupled BLG state, there exist two Rashba-coupling modes in terms of spin chirality: isochiral coupling ($\nu=1$) and opposite-chiral coupling [$\nu=-1$, see Fig.~1(c)].

We are mainly concerned with the physical effects induced by the sign change of the interlayer Rashba polarization parameter $\nu$. We summarize the case of uniform-RSOC ($\nu=1$) as follows. The lowest-energy four subbands are expressed as \cite{ZhaiJin} $\varepsilon_{\alpha\beta}(\bm p)=\alpha \upsilon|\bm p|[(\lambda^2+\epsilon^2)^{1/2}-\beta\lambda]/2$ with $\alpha,\beta=\pm1$. The average spin is solved as
$\langle{\bm s}\rangle=\beta{\rm sgn}(\lambda)\epsilon(\lambda^2+\epsilon^2)^{-1/2}(\hat{{\bm e}}_{\bm p}\times\hat{\bm z})$, where the orientation of spin chirality depends on the sign of the index $\beta$. For the case of LO-RSOC ($\nu=-1$), we surprisingly find that the eight-band Hamiltonian~(\ref{H_EB}) always has the following four eigenvalues
\begin{equation}\label{LO-Band}
\varepsilon_0^{\nu=-1}=\pm\frac{1}{2}\left(\Gamma\pm\sqrt{4\upsilon^2p^2+\Gamma^2}\right),
\end{equation}
where $\Gamma=(4\lambda^2+\gamma^2)^{1/2}$ holds. Consequently, no spin splitting occurs, and spin chirality disappears ($\langle{\bm s}\rangle=0$).

\begin{figure}
\centerline{\includegraphics[width=6.5cm]{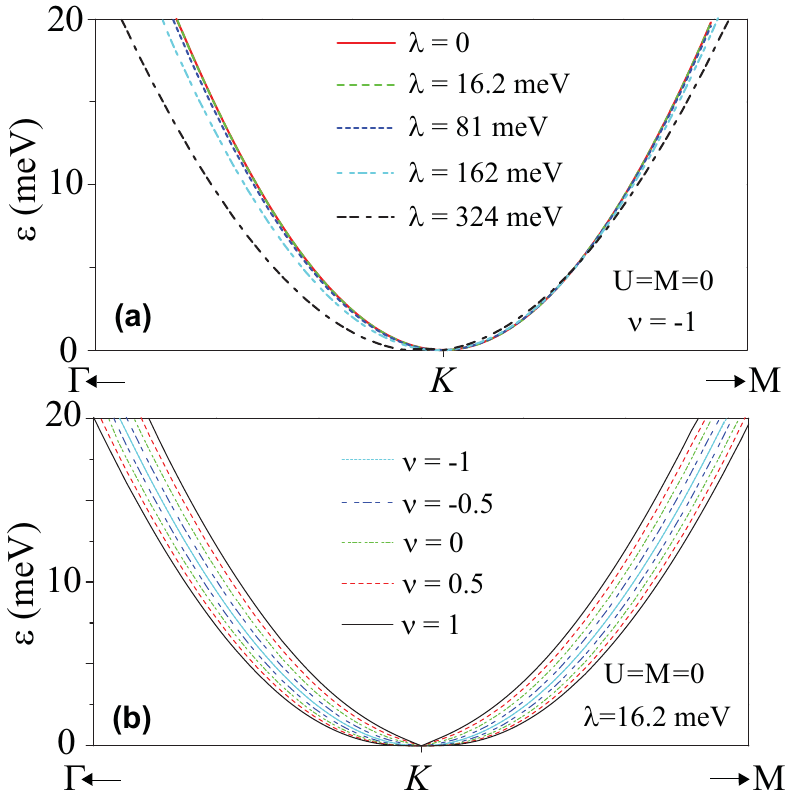}} \caption{(a) Band structures for $\lambda$ increasing from 0 to 324~meV through 16.2~meV, 81~meV and 162~meV. (b) Band structures for $\nu$ changing from -1 to 1 through -0.5, 0, 0.5. We set $U=M=0$ in both (a) and (b), and in this case, two degenerate valleys are associated with each other by $\cal T$ symmetry.}
\end{figure}

We deeply argue the striking phenomenon of chiral loss induced by LO-RSOC in Fig.~1(c). As is understood, it is the interlayer vdW coupling that mixes opposite spin chiralities on opposite layers and enables the chiral loss in total. This raises a problem that it is hard to distinguish the band difference between $\nu=-1$ case and no-Rashba case, as shown in Figs.~1(d) and 1(e), or rather, it is difficult to judge whether the LO-RSOC is present. Note that $\lambda$ here is on the order of 10~meV, which is readily available in experiment \cite{BalKok,HanKaw,AvsOch,MarVar,ManKoo}. For a giant $\lambda$ comparable to $\gamma$ (more than 0.1~eV), the band slope gets visibly lower, as shown in Fig.~2(a). Moreover, we plot the band structures for $\nu$ changing from -1 to 1 through -0.5, 0, 0.5 in Fig.~2(b). As is seen, the band spin degeneracy is opened as long as $\nu\neq-1$.

\section{Field-induced symmetry breaking}

We naturally ask ``Are field-induced symmetry breaking helpful to identify the presence of LO-RSOC?" In terms of actual experimental realizability, $H_U$ in Hamiltonian~(\ref{H_EB}) is feasible in dual-gated device, while $H_\lambda$ and $H_M$ are simultaneously inducible by contacting graphene with, for example, Cr$_2$Ge$_2$Te$_6$ under pressure \cite{ZhangZhao} or magnetic layers of Co (Ni) \cite{PerMed}.

Under the low-energy approximation ($\varepsilon<\gamma$), the eight-band model Hamiltonian~(\ref{H_EB}) can be further reduced to the four-band form [see Eq.~(\ref{H_FB})] that captures the lowest-energy four bands closest to the Fermi energy, by employing van Vleck's perturbation theory \cite{vanVleck}. The processing method is as follows.

Taking $\psi=\{\psi_{A_2\uparrow},\psi_{A_2\downarrow},\psi_{B_1\uparrow},
\psi_{B_1\downarrow},\psi_{A_1\uparrow},\psi_{A_1\downarrow},\psi_{B_2\uparrow},
\psi_{B_2\downarrow}\}$ as the atomic basis set, the low-energy effective Hamiltonian~(\ref{H_EB}) in the main text is rewritten as
\begin{equation}\label{H0W}
\begin{split}
H=H_0+W,~H_0=
\left[\begin{matrix}
H_+&0\\
0&H_-\\
\end{matrix}\right],~
W=
\left[\begin{matrix}
0&H_s\\
H_s^\dag&0\\
\end{matrix}\right],\\
\end{split}
\end{equation}
where the diagonal matrices $H_\pm$ read
\begin{equation*}
\begin{split}
H_\pm=&\left[
\begin{matrix}
\mp U+M&0,&0,&0\\
0,&\mp U-M,&0,&0\\
0,&0,&\pm U+M,&0\\
0,&0,&0,&\pm U-M
\end{matrix}
\right],\\
\end{split}
\end{equation*}
and the valley-dependent matrix $H_s$ is described by
\begin{equation*}
\begin{split}
H_s=&\left[
\begin{matrix}
0&0,&\gamma\pi,&\frac{\xi-1}{2}i\nu\lambda\\
0,&0,&\frac{\xi+1}{2}i\nu\lambda,&\gamma\pi\\
\gamma\pi^\dag,&-\frac{\xi+1}{2}i\lambda,&0,&0\\
\frac{1-\xi}{2}i\lambda,&\gamma\pi^\dag,&0,&0
\end{matrix}
\right],
\end{split}
\end{equation*}
where $\pi=\upsilon(p_x-i\xi p_y)$ is defined.

By using matrix diagonalization, the eigenvalues of $H_0$ are solved as
\begin{equation}\label{EVs}
\begin{split}
\varepsilon_{1,2}^0=&\mp M-\sqrt{\gamma^2+U^2},\\
\varepsilon_{3,4}^0=&-U\mp M,\\
\varepsilon_{5,6}^0=&U\pm M,\\
\varepsilon_{7,8}^0=&\mp M+\sqrt{\gamma^2+U^2},\\
\end{split}
\end{equation}
corresponding to the eigenvector $|\Psi^0\rangle=(|\psi_1^0\rangle,|\psi_2^0\rangle,\ldots,|\psi_8^0\rangle)$ written as
\begin{equation}
|\Psi^0\rangle=\left[\begin{matrix}
0&0&0&1&0&0&0&0\\
0&0&1&0&0&0&0&0\\
0&0&0&0&1&0&0&0\\
0&0&0&0&0&1&0&0\\
0&-\zeta&0&0&0&0&0&\vartheta\\
-\zeta&0&0&0&0&0&\vartheta&0\\
0&\vartheta&0&0&0&0&0&\zeta\\
\vartheta&0&0&0&0&0&\zeta&0\\
\end{matrix}
\right],
\end{equation}
where $\vartheta=\cos(\varphi/2)$, $\zeta=\sin(\varphi/2)$ and $\tan\varphi=\gamma/U$.
We divide the eight eigenvalues into two groups in terms of energy,  $\varepsilon_{ia}^0\in\{\varepsilon_1^0,\varepsilon_2^0,
\varepsilon_7^0,\varepsilon_8^0\}$ and $\varepsilon_{jb}^0\in\{\varepsilon_2^0,\varepsilon_3^0,\varepsilon_4^0,
\varepsilon_5^0\}$, satisfying $|\varepsilon_{ia}^0-\varepsilon_{ja}^0|\sim|\varepsilon_{ib}^0-\varepsilon_{jb}^0|
\ll|\varepsilon_{ia}^0-\varepsilon_{jb}^0|\sim\gamma$.
The low-energy Hamiltonian for BLG is thus achievable through the unitary transformation $\tilde{H}={\rm e}^{iS}H{\rm e}^{-iS}$, where the $S$ matrix elements are given by
\begin{equation}
\begin{split}
S_{ml}=&\frac{iW_{ml}}{\varepsilon_l^0-\varepsilon_m^0}+i\sum_{m'}\frac{W_{mm'}W_{m'l}}
{(\varepsilon_l^0-\varepsilon_m^0)(\varepsilon_l^0-\varepsilon_m'^0)}\\
&+i\sum_{l'}\frac{W_{ml'}W_{l'l}}
{(\varepsilon_l^0-\varepsilon_m^0)(\varepsilon_l^0-\varepsilon_{l'}^0)}.
\end{split}
\end{equation}
Herein, $S=S^\dag$, $W_{ml}=\langle\psi^0_m|W|\psi^0_l\rangle$, $m,m'\in\{3,4,5,6\}$ and $l,l'\in\{1,2,7,8\}$ hold.
The low-energy matrix elements of the effective Hamiltonian (up to second order in $1/\gamma$) are determined by
\begin{equation}\label{Hmn}
\begin{split}
H_{mm'}=&\varepsilon_m^0\delta_{mm'}+W_{mm'}\\
&+\frac{1}{2}\sum_l
W_{ml}W_{lm'}\left(\frac{1}{\varepsilon_m^0-\varepsilon_l^0}
+\frac{1}{\varepsilon_{m'}^0-\varepsilon_l^0}\right)+{\cal O}(2),
\end{split}
\end{equation}
with $H_{mm'}=(H_{m'm})^\dag$. By using Eq.~(\ref{Hmn}), we derive the effective Hamiltonian~(4) in the main text.

According to Eq.~(\ref{Hmn}), under $\{\varepsilon,M,U\}<\gamma$ and $\{\lambda^2/\gamma^2,\lambda M/\gamma^2,\lambda U/\gamma^2\}\rightarrow0$, we obtain the lowest-energy four-band Hamiltonian as
\begin{equation}\label{H_FB}
\begin{split}H_{\rm eff}=&H^{(0)}+H^{(1)}+H^{(2)}+{\cal{O}}(1/\gamma^3),\\
H^{(0)}=&-\sigma_z(U{\bm I_s}+\xi Ms_z),\\
H^{(1)}=&-\frac{1}{\gamma}
\left(\begin{array}{ccc}
0&(\pi^\dag)^2\\
\pi^2&0\\
\end{array}\right) s_x
+\Theta\frac{i(1+\nu)\lambda}{\gamma}s_\pi\sigma_+,\\
H^{(2)}=&\frac{2U}{\gamma^2}
\left(\begin{array}{ccc}
\pi^\dag\pi&0\\
0&-\pi^\dag\pi\\
\end{array}\right){\bm I_s},
\end{split}
\end{equation}
in the atomic basis set $\psi=\{\psi_{A_2\downarrow},\psi_{A_2\uparrow},\psi_{B_1\uparrow},
\psi_{B_1\downarrow}\}$ for valley $K$ and  $\psi=\{\psi_{A_2\uparrow},\psi_{A_2\downarrow},
\psi_{B_1\downarrow},\psi_{B_1\uparrow}\}$ for valley $K'$.
Here, we define $s_\pm=(s_0\pm s_z)/2$, $\sigma_\pm=(\sigma_0\pm\sigma_z)/2$,
\begin{equation}
s_\pi=\left(\begin{array}{ccc}
0&\pi^\dag\\
-\pi&0\\
\end{array}\right),~~~\Theta=\frac{1}{2}\left(1+\frac{\gamma^2}{\gamma^2+\Delta}\right)
\end{equation}
to shorten notation, with $\Delta=4M(U-M)$. Note that $s_\pi$ is in the $\bm s$ space, $\Theta$ is the dimensionless factor renormalized by $U$ and $M$ ($\Theta=1$ for $U=M=0$), and the ${\bm \sigma}$ space here refers to the A$_2$ and B$_1$ sublattices [different from that in Hamiltonian~(\ref{H_EB})]. Judged from Hamiltonian~(\ref{H_FB}), the LO-RSOC leads to the factor $1+\nu=0$, which is responsible for the chiral loss in Eq.~(\ref{LO-Band}), reflecting no spin polarization due to the cancellation of both layers.

\begin{figure}
\centerline{\includegraphics[width=8.5cm]{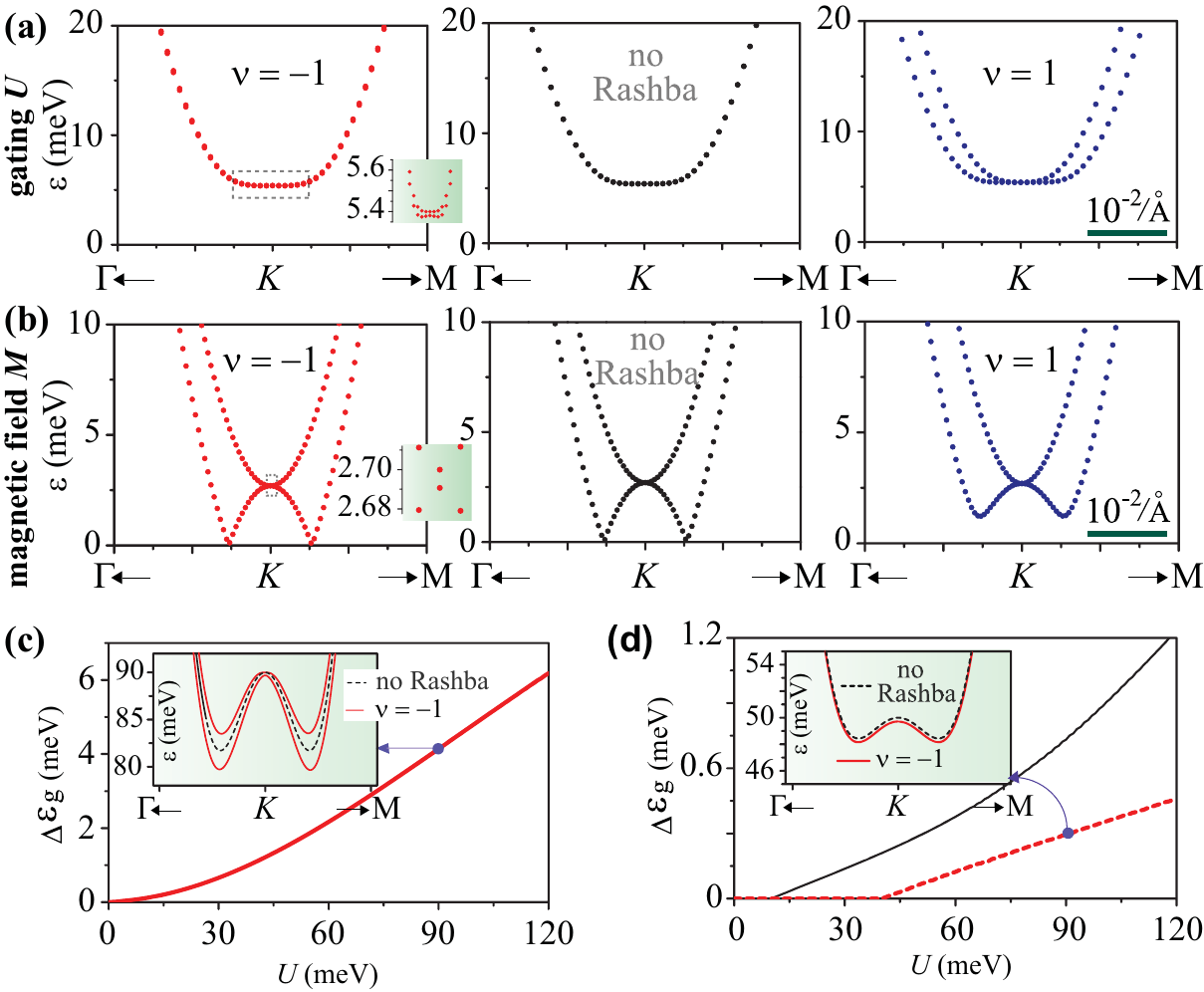}} \caption{Band structures modulated by (a) gating $U=5.4$~meV and (b) magnetic field $M=2.7$~meV on basis of  Figs.~1(b)-1(d). The insets enlarge the dotted-box regions. (c) and (d) Dependence of bandgap difference between no-Rashba case and $\nu=-1$ case on $U$, for (c) $M=0$ and (d) $M\neq0$ [10~meV (solid line) and 40~meV (dashed line)]. The insets show the band details.}
\end{figure}

In Figs.~3(a) and 3(b), we illustrate the influence of $U=5.4$~meV and $M=2.7$~meV, respectively, on the conduction bands in Figs.~1(d)-1(f). Our results indicate that it is still hard to observe the band difference between no-Rashba case and $\nu=-1$ case, because the spin splitting induced by $U$ ($M$) at $K$ is only about 37~$\mu{\rm eV}$ (19~$\mu{\rm eV}$), which is indeed negligible as expected from Hamiltonian~(\ref{H_FB}).

For larger values of $U$ or $M$, we need to add the other perturbation contributions $\delta H^{(2)}$ to $H^{(2)}$ as follows
\begin{equation}\label{dH}
\begin{split}
\delta H^{(2)}=&\frac{\lambda^2}{\gamma^2}\left(\begin{array}{ccc}
\nu^2{\cal J}&0\\0&-2U\\
\end{array}\right)s_++\frac{i\lambda}{\gamma^2}
(\nu{\cal J}\sigma_+ s_\pi^*+2U\sigma_-s_\pi),
\end{split}
\end{equation}
where ${\cal J}=U+(2\Theta-1)(U-M)$ holds. In Figs.~3(c) and 3(d), we further plot the bandgap difference $\Delta\varepsilon_g$ (between no-Rashba case and $\nu=-1$ case) modulated by $U$ for $M=0$ and $M\neq0$, respectively. It is shown that the nonzero bandgap increases as $U$ increases [$\varepsilon_g=0$ is always valid for $\Delta\varepsilon_g=0$ in Fig.~3(d)]. Taking $U=120$~meV for example, $\Delta\varepsilon_g$ is about 6~meV, 1.2~meV, 0.45~meV for $M=0$, 10~meV and 40~meV, respectively. Therefore, increasing $U$ is helpful to identify the LO-RSOC by enhancing spin splitting. By contrast, increasing $M$ lowers the splitting. The second-order perturbation effect reflected by Eq.~(\ref{dH}) in the BLG-based LO Rashba system reveals the weak sensitivity of Rashba splitting on the electrically or magnetically induced symmetry breaking, in contrast to the strong sensitivity of that in other known Rashba bilayers to even weak symmetry-breaking perturbations \cite{DasBal,RajBan,VolLoss,Atkin,LuSen,LinWang,YuanLiu,XuXia,ChenKanou}. This weak sensitivity should be attributed to the specific vdW-coupled structure of BLG, and supports BLG to hold extremely-stable band topology in a relatively complex double-interface problem \cite{IsLew}.

\section{Topological transition}

Now, it is necessary to clarify what happens when the amplitude homogeneity of RSOC between two monolayers is broken, corresponding to $|\nu|\neq1$ in Hamiltonian~(\ref{H_EB}). In practice, adjusting the concentration of adatoms \cite{BalKok} from one side (top or bottom) of BLG or fabricating asymmetric vertical heterostructures are feasible to induce the interlayer Rashba inhomogeneity. The existing theoretical data \cite{ZolFab} also suggests that twisting the angle between graphene and its proximity material may alter the value of $\nu$ by breaking the heterostructure symmetry.

In the absence of $H_U$ and $H_M$, the RSOC itself in Hamiltonian~(\ref{H_EB}) does not open a bandgap (independent of $\nu$), determined by Eq.~(\ref{LO-Band}). The system is not a topological insulator but a semimetal. Nevertheless, the Rashba system exhibits the intriguing Fermi-loop topology, characterized by the Berry's phase or geometric phase \cite{Zak}, defined by $\gamma_n=\oint_Cd{\bm p}\cdot{\cal A}_n(\bm p)$, where ${\cal A}_n(\bm p)=\langle\psi_n(\bm p)|i{\bm \nabla}_{\bm p}|\psi_n(\bm p)\rangle$ is the Berry connection for the wave function $\psi_n(\bm p)$ in the $n$-th subband. The wave function of Hamiltonian~(\ref{H_FB}) is solved as $\psi_{n\supseteq\{\alpha,\beta\}}^T=(-\alpha\zeta_{-\beta}{\rm e}^{-i\xi\phi},-i\alpha\beta\zeta_\beta{\rm e}^{-2i\xi\phi},
i\beta\zeta_{-\beta},\zeta_\beta{\rm e}^{i\xi\phi})/\sqrt2$, corresponding to the dispersion
$\varepsilon_{\alpha\beta}(\bm p)=\alpha\upsilon|\bm p|[2(\lambda^2+\epsilon^2)^{1/2}-\beta(1+\nu)\lambda]/4$. Here we have  $\alpha,\beta=\pm1$, $\tan\phi=p_y/p_x$, $\zeta_\beta=\delta_{\beta,1}\cos(\theta/2)+\delta_{\beta,-1}\sin(\theta/2)$ and $\tan\theta=2\upsilon|\bm p|/[(1+\nu)\lambda]$, with $\delta_{ij}$ denoting the Kronecker delta function.
Strikingly, $\gamma_n$ is derived as
\begin{equation}\label{BP}
\gamma_n=
2\pi~(\nu=-1),~~
\pi~(\nu\neq-1).
\end{equation}
Note that we ignore the sign of $\gamma_n$, for which $\pm2\pi$ $(\pm\pi)$ are equivalent because the phase period is $2\pi$. As a result, a sharp transition of the Berry's phase appears when $\nu$ deviates from -1. This transition is suggested to be detected by the contrasting conductance through an $np$ junction based on the fact that $\gamma_n=\pi$ supports Klein tunneling but $\gamma_n=2\pi$ does not \cite{Neto}. In experiment, it requires the adequate low-temperature condition to avoid the interband scattering and ensure the ballistic transport. Notably, the local gauge-invariant quantity defined as \cite{XiaoChang} ${\bm \Omega}_n(\bm p)={\bm \nabla}_{\bm p}\times{\cal A}_n(\bm p)$ (Berry curvature) is always zero for the gapless and $\cal T$-symmetry case here.

\begin{figure}
\centerline{\includegraphics[width=8.5cm]{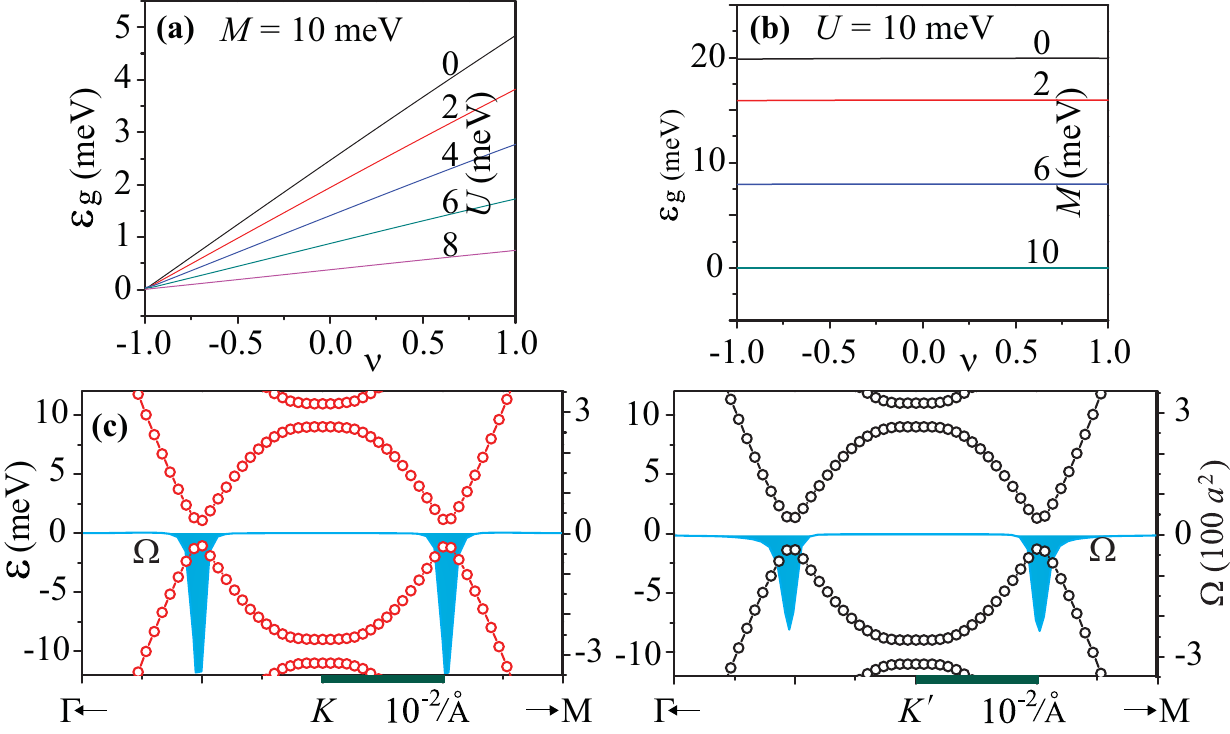}} \caption{(a) and (b) Bandgap versus polarization parameter $\nu$ by fixing $M=10$~meV and varying $U$ in (a) and by fixing $U=10$~meV and varying $M$ in (b). (c) Band structure and Berry curvature ($\Omega$) near two valleys for $(\nu,M,U)=$(-0.1, 10~meV, 1~meV). In the unit notation for $\Omega$, $a=2.46~{\AA}$ is the lattice constance of BLG.}
\end{figure}

When $H_U$ or $H_M$ are present, the Rashba system is usually gapped \cite{MirSch,QiaoLi,QiaoTse}. We show the bandgap as a function of $\nu$ in Figs.~4(a) and 4(b), where $M=10$~meV and $U=10$~meV are fixed, respectively. The results indicate that, only for $M>U$, $\nu$ changes the bandgap. As $\nu$ gets closer to 1 [see Fig.~4(a)], the bandgap becomes larger. For the gapped system, $\gamma_n$ is a variable that depends on momentum and thus no longer provides a good topology description. The invariant to characterize the band topology here is the Chern number determined by ${\cal C}=\sum_{n\in{\rm VB}}\int_{\rm BZ}{\bm \Omega}_n(\bm p)d^2p/(2\pi)^2$, where VB (BZ) denotes the valence bands (Brillouin zone). For $U^2<M^2$, combined with the condition
$\{M^2,\lambda^2\}\ll\gamma^2$ readily achievable in experiment \cite{KriGol,ManKoo,AvsOch,BalKok,MarVar,HanKaw}, it satisfies
\begin{equation}\label{Chern}
{\cal C}=
0~(\nu=-1),~-2{\rm sgn}(M)~(\nu\neq-1).
\end{equation}
For $U^2>M^2$, we always have ${\cal C}=0$, but the system is a quantum valley Hall insulator \cite{QiaoLi}, independent of $\nu$, because $\nu$ does not alter the bangap, as shown in Fig.~4(b).

We plot the band structure and Berry curvature for a topological insulating state with a set of parameters $(\nu,M,U)=$(-0.1, 10~meV, 1~meV). It is seen that valley degeneracy is broken, reflected by the differences of band structure and Berry curvature near two valleys. Nevertheless, the Chern number ${\cal C}=-2$ is contributed equally by two valleys.

Moreover, it should be noted that Eqs.~(\ref{BP}) and (\ref{Chern}), which reveal the $\nu$-related sharp topological transitions, are our significant results for the BLG-based LO-RSOC system. No evidence of these sharp topological transitions has been found in other non-graphene Rashba bilayers \cite{DasBal,RajBan,VolLoss,Atkin,LuSen,LinWang,YuanLiu,XuXia,ChenKanou}.

\section{LO-RSOC confirmed by ab-initio calculations}

Beyond the phenomenological Hamiltonian~(\ref{H_EB}), we further show a concrete LO-RSOC system that is BLG encapsulated by two monolayers of Au, as shown in the left panel of Fig.~4(a), where the optimized stable structure is ${\cal I}$ symmetry. We employ the standard {\it ab initio} calculations that are performed in MLG--Au interface \cite{KriGol}, where the Rashba strength $\lambda$ depends strongly on the graphene-Au distance $d_{\rm G-Au}$ and is negligible for $d_{\rm G-Au}>4.2~{\AA}$. We obtain the optimized interlayer distance $d_{\rm G-Au}=3.21~{\AA}$.

\begin{figure}
\centerline{\includegraphics[width=8.0cm]{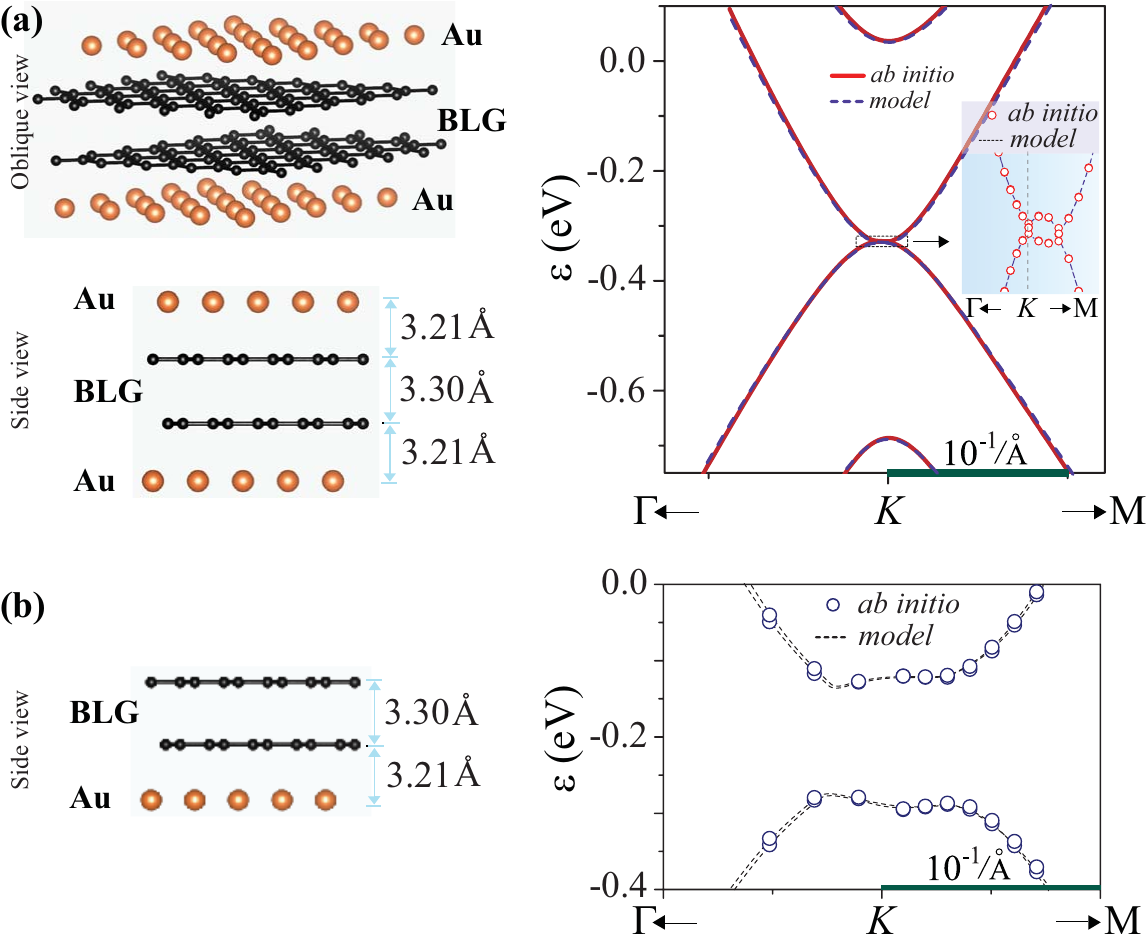}} \caption{Real-space lattice structure (left) and band structure (right) for (a) BLG encapsulated by two single layers of Au and (b) BLG in proximity with single layer of Au (for comparison). The optimized layer distances in (a) are marked. To obtain the value of $\lambda$ in (a), two layer distances in (b) are artificially set to be the same with (a). The band structures are obtained by {\it ab inito} calculations and fitted by the model. The inset in (a) enlarges the region of trigonal warping.}
\end{figure}

The calculated band structure in the right panel of Fig.~5(a) shows that no spin splitting happens, as expected from our prediction of the LO-RSOC in Hamiltonian~(\ref{H_EB}). It is shown that BLG becomes electron doping (chemical potential $\mu=-0.329$~eV) due to the $\pi$--$d$ orbital hybridization between graphene and Au. This means, the direction of charge transfer at each hetero-interface is from Au to graphene. Because the top Au--graphene interface and the bottom graphene--Au interface are ${\cal I}$ symmetry, the interfacial potential gradient along the $z$ axis to induce the LO-RSOC \cite{BerLuc} is opposite in sign on opposite layers of BLG. The hyperfine band structure near $K$ (see the inset) reveals that the trigonal warping effect ($|\varepsilon-\mu|<1$~meV) does not open the spin degeneracy.

Now, how to determine the value of $\lambda$ in Fig.~5(a) is still a question, because the band structure can be well fitted by the established model ($t=2.689$~eV and $\gamma=0.386$~eV) even without $\lambda$, whereas $\lambda$ is not negligible in each monolayer graphene. More strictly, there still need additional interlayer hopping parameters to achieve a better fit, including the nearest-neighbor hopping energy $\gamma'=0.296$~eV between sublattice B$_1$ and A$_2$, and the nearest-neighbor hopping energy $\gamma''=0.0382$~eV between sublattice A$_1$ (B$_1$) and A$_2$ (B$_2$). To obtain a relatively-accurate value of $\lambda$, we further perform the {\it ab initio} calculations for the BLG by proximity with monolayer Au in Fig.~5(b), where two layer distances are manually set to be the same with Fig.~5(a). It is shown than a gap is opened, and spin splitting appears. The parameters $\lambda=16.2$~meV, $\nu=0$ and $U=0.086$~eV in model~(\ref{H_EB}) are suitable to fit the band data in Fig.~5(b). By comparing Figs.~5(a) and 5(b), we conclude that $\lambda=16.2$~meV and $\nu=-1$ in Fig.~5(a) hold.

Undoubtedly, the spacial-distribution and concentration of Au atoms or the additional use of magnetic Ni substrate have obvious influence on the change of structural symmetry and band structure \cite{MarVar,KriGol}. Au layer in Fig.~4(a) only provides a simple example and is actually optional to induce the Rashba effect. Essentially, our model Hamiltonian~(\ref{H_EB}) captures the main Rashba physics of all the possible doubly-proximitized BLG systems by adjusting $\nu$, $U$ and $M$, and sometimes, by adding other nessessary interactions such as staggered sublattice potential, Ising or Kane-Mele SOCs \cite{LopCol,WangChe,ZhaiXu,BalKok,IsLew,AlsAsm}.

\section{Conclusions}

We have revealed that the ${\cal I}$ symmetry in BLG allows the presence of LO-RSOC, which results in the loss of spin chirality in the momentum space and is identifiable by ${\cal I}$ or ${\cal T}$ symmetry breaking through inducing spin splitting or driving the transition of band topology. These nontrivial results are fundamental to understanding the Rashba physics in all the possible 2D layered structures which are doubly-proximitized from both the top and bottom sides, and pave the way to developing 2D spintronics by fully activating the dimension of layer besides spin.

\section*{Acknowledgments}

This work was supported by the NSFC with Grant Nos. 12074193 and 61874057. We
thank D. Marchenko and Y. M. Blanter for helpful discussions. Thank Y. M. Blanter for his efforts in revising the language.

\end{document}